\documentclass[letterpaper,10pt,conference]{ieeeconf}
\IEEEoverridecommandlockouts
\usepackage{cite}
\usepackage{amsmath,amssymb,amsfonts}
\usepackage{graphicx}
\usepackage{tabularx}
\usepackage{xcolor}
\usepackage{subfigure}
\usepackage{epstopdf}

\title{\LARGE \bf Neural Signatures of Motor Skill in the Resting Brain}

\author{Ozan \"{O}zdenizci$^{1,2,*}$, Timm Meyer$^{3}$, Felix Wichmann$^{4}$, Jan Peters$^{5}$, Bernhard Sch\"{o}lkopf$^{3}$,\\M\"{u}jdat \c{C}etin$^{2,6}$ and Moritz Grosse-Wentrup$^{3,7}$
\thanks{$^{1}$Department of Electrical and Computer Engineering, Northeastern University, Boston, MA, United States.}%
\thanks{$^{2}$Signal Processing and Information Systems Laboratory, Faculty of Engineering and Natural Sciences, Sabanc{\i} University, Istanbul, Turkey.}%
\thanks{$^{3}$Max Planck Institute for Intelligent Systems, T\"{u}bingen, Germany.}%
\thanks{$^{4}$Faculty of Science, University of T{\"u}bingen and Bernstein Center for Computational Neuroscience T{\"u}bingen, T{\"u}bingen, Germany.}%
\thanks{$^{5}$FG Intelligente Autonome Systeme, Technische Universit\"at Darmstadt, Darmstadt, Germany.}%
\thanks{$^{6}$Department of Electrical and Computer Engineering, University of Rochester, Rochester, NY, United States.}%
\thanks{$^{7}$Research Group Neuroinformatics, Faculty of Computer Science, University of Vienna, Vienna, Austria.}%
\thanks{$^{*}$Corresponding author: {\tt\small oozdenizci@ece.neu.edu}}%
}

\begin{document}

\maketitle
\thispagestyle{empty}
\pagestyle{empty}


\begin{abstract}
Stroke-induced disturbances of large-scale cortical networks are known to be associated with the extent of motor deficits. We argue that identifying brain networks representative of motor behavior in the resting brain would provide significant insights for current neurorehabilitation approaches. Particularly, we aim to investigate the global configuration of brain rhythms and their relation to motor skill, instead of learning performance as broadly studied. We empirically approach this problem by conducting a three-dimensional physical space visuomotor learning experiment during electroencephalographic (EEG) data recordings with thirty-seven healthy participants. We demonstrate that across-subjects variations in average movement smoothness as the quantified measure of subjects' motor skills can be predicted from the global configuration of resting-state EEG alpha-rhythms (8--14 Hz) recorded prior to the experiment. Importantly, this neural signature of motor skill was found to be orthogonal to (independent of) task- as well as to learning-related changes in alpha-rhythms, which we interpret as an organizing principle of the brain. We argue that disturbances of such configurations in the brain may contribute to motor deficits in stroke, and that reconfiguring stroke patients' brain rhythms by neurofeedback may enhance post-stroke neurorehabilitation.
\end{abstract}

\section{Introduction}

Motor learning is a major component in the field of robotic neurorehabilitation as it is known to be analogous to post-stroke recovery in humans \cite{Reinkensmeyer:2004,Krakauer:2006}. Neuroimaging introduces various evidence on distinct large-scale cortical networks (LSCNs) \cite{Bressler:2010} being involved in different stages of motor learning \cite{Hikosaka:2002,Grafton:2007}. Furthermore, post-stroke disturbances of LSCNs, measured by functional magnetic resonance imaging (fMRI), have been shown to correlate with the severity of motor deficits \cite{Sharma:2009,Inman:2012,Wang:2014,Thiel:2015}. Because stroke recovery is a form of motor learning and motor learning involves reconfigurations of LSCNs \cite{Bassett:2011}, we hypothesize that stroke-induced disturbances of LSCNs do not only reflect, but actively contribute to the severity of motor deficits. Hence, we argue that reconfiguration of LSCNs representative of motor behavior in the resting brain could provide significant support for current post-stroke neurorehabilitation protocols.

Over the last two decades, research on electroencephalogram (EEG) based brain-computer interfaces (BCIs) generated significant interest in post-stroke neurorehabilitation, considering that EEG is the only neuroimaging modality that can easily be integrated into a rehabilitation robot \cite{Birbaumer:2007}. In that context, it has been argued that synchronization of movement intent, as decoded by monitoring patients' sensorimotor rhythms (SMRs), with congruent haptic feedback, delivered through robotic-based movement support  \cite{Gomez:2011,Ramos:2012,Sarac:2013,Ang:2014,Ang:2015}, may reinforce neural plasticity and enhance motor recovery \cite{Grosse-Wentrup:2011}. To date, two studies have reported empirical evidence on beneficial effects of BCI-based rehabilitation training in patients with subacute \cite{Pichiorri:2015} and with chronic stroke \cite{Ramos:2013}. Despite this recent progress, the extent of brain activities that are considered, which are restricted to SMRs, can be a confounding factor for further progress. Instead of focusing exclusively on sensorimotor activity of the brain, we argue that BCI-based neurorehabilitation systems should further consider EEG signatures of LSCNs in the resting brain \cite{Mantini:2007}, that are related to human motor control.

Several EEG decoding studies aim to investigate global neural correlates of human motor behavior by quantifying motor learning performance with measures of adaptation or learning rates \cite{Wu:2014,Ozdenizci:2017,Meyer:2014,Manuel:2018}. However, the significant dissociation between motor skill and motor learning performance, where skill shapes movements and motor control during learning \cite{Krakauer:2011,Schmidt:2018}, is yet to be explored. In particular, the concept of movement smoothness, which was introduced by minimum-jerk models in motor learning \cite{Flash:1985} and extensively studied in post-stroke recovery processes \cite{Krebs:1999,Rohrer:2002}, is considered to be a viable index of motor skill, as recently used in fMRI neuroimaging studies \cite{Lefebvre:2012,Sosnik:2014}.

In this study, we investigate how motor skills of healthy subjects in a three-dimensional physical space visuomotor learning task is represented by a global configuration of their EEG rhythms recorded prior to the task at rest. We show that across-subjects variations in movement smoothness can be predicted with a neural signature of motor skill consisting of a network of resting-state brain rhythms. Furthermore, we observe the dynamics of this neural signature to be independent of motor planning, execution, and the learning process. In the context of BCI-based neurorehabilitation, we hypothesize that disturbances of such global configurations of brain rhythms contribute to motor deficits in stroke, and that reconfiguring patients' brain rhythms by BCI-based neurofeedback may potentially enhance motor recovery.

\section{Materials and Methods}

\subsection{Participants}

Thirty-seven healthy right-handed male subjects (mean age $27.8 \pm 7.1$), recruited from the T\"ubingen area, participated in the study. By restricting the participants to male subjects, we minimized variance in muscle strength. All subjects performed the visuomotor learning task with their right arms. All subjects were naive to the task. Before conducting the study, an experimenter explained the experimental procedure to the subjects and obtained their informed consent. The design and experimental setup of this research study were approved by the ethics committee of the Max Planck Society. The experimental data we collected for this study is publicly available at: https://doi.org/10.5281/zenodo.1434948.

\subsection{Study Design}

\begin{figure}[t!]
    \centering
	\subfigure[]{\includegraphics[width=0.22\textwidth]{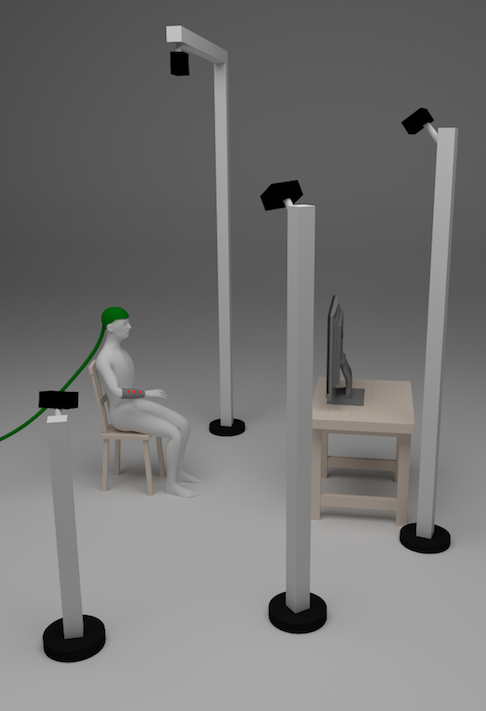}\label{fig:visSetup}}
	\hspace{0.02\textwidth}
	\subfigure[]{\includegraphics[width=0.22\textwidth]{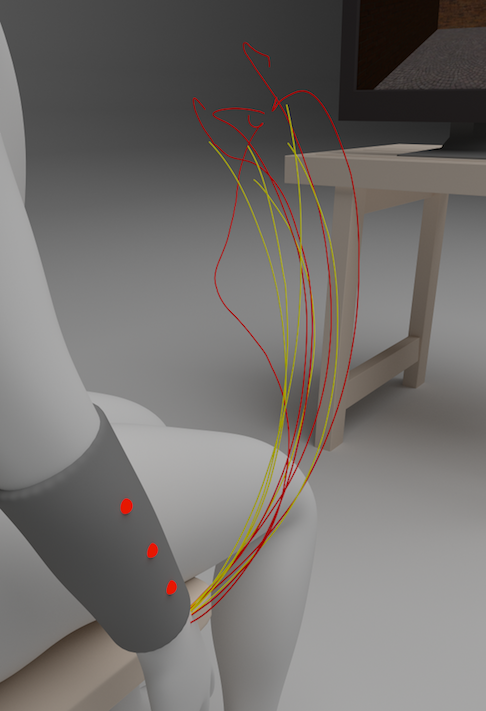}\label{fig:visPaths}}
	\caption{(a) Overview of the experimental setup. (b) Sample reaching movements from one participant's kinematic data recordings. Yellow trajectories demonstrate low jerk with high score, and red trajectories demonstrate high jerk with low score reaching movements.}\vspace{-0.4cm}
	\label{fig1}
\end{figure}

During the experiments, subjects were comfortably placed approximately 1.5 meters in front of a computer screen displaying a three-dimensional virtual reality scene. Their reaching movements were captured by an Impulse X2 Motion Capture System (PhaseSpace, San Leandro, CA, U.S.) at a frame rate of 960 Hz (see Fig.~\ref{fig:visSetup}). Motion capturing was realized with three infra-red LED markers placed alongside a forearm sleeve, worn on the right arm of the subject. These LEDs defined a rigid body, henceforth referred to as an \textit{end-effector}, captured by four cameras placed around the physical reaching movement area.

The task in each trial was to move the end-effector to a target location in the three-dimensional space, and then back to the starting position, while the entire movement is continuously monitored on the computer screen in real time. At the end of each trial, subjects were provided with visual feedback on their motor performance, represented by a trial score (i.e., \textit{rating}) within a range of 0--100, based on the smoothness of movement (see Section~\ref{sec:narj} for a detailed description). Subjects who produced higher-rated movements received additional monetary compensation, incentivizing them to increase their scores throughout the experiment. However, the subjects were not informed of the precise link between the rating and their movement characteristics. Throughout the experiments, each subject performed 250 trials, divided into five blocks of 50 trials, separated by one to two minute intermissions. In the third and fourth blocks, a 1-kg weight was attached to the right hand of the subject. For this study, we are only interested in the analysis of the first 100 reaching movements (i.e., first two experimental blocks) that each subject performed, which does not involve external manipulations or disturbances of movements. We further observed that successful visuomotor learning was accomplished within these 100 reaching movements, and performance converged to a steady-state behavior by the end. Rest of the data were subject to analysis for another study.

\begin{figure*}[h!]
    \centering
	\subfigure[]{\includegraphics[width=0.24\textwidth]{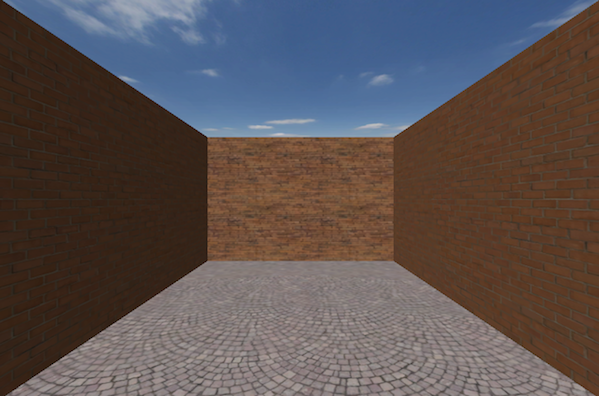}\label{fig:visBaseline}}
	\hspace{0.01\textwidth}
	\subfigure[]{\includegraphics[width=0.24\textwidth]{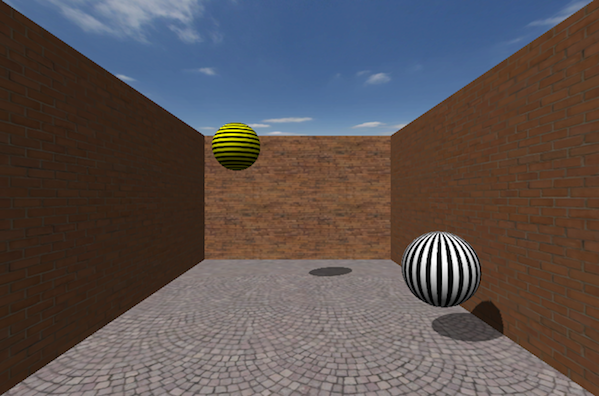}\label{fig:visPlanning}}
	\hspace{0.01\textwidth}
	\subfigure[]{\includegraphics[width=0.24\textwidth]{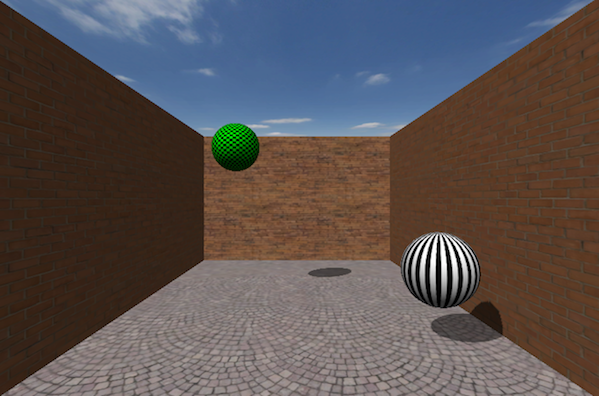}\label{fig:visReaching}}
	\hspace{0.01\textwidth}
	\subfigure[]{\includegraphics[width=0.24\textwidth]{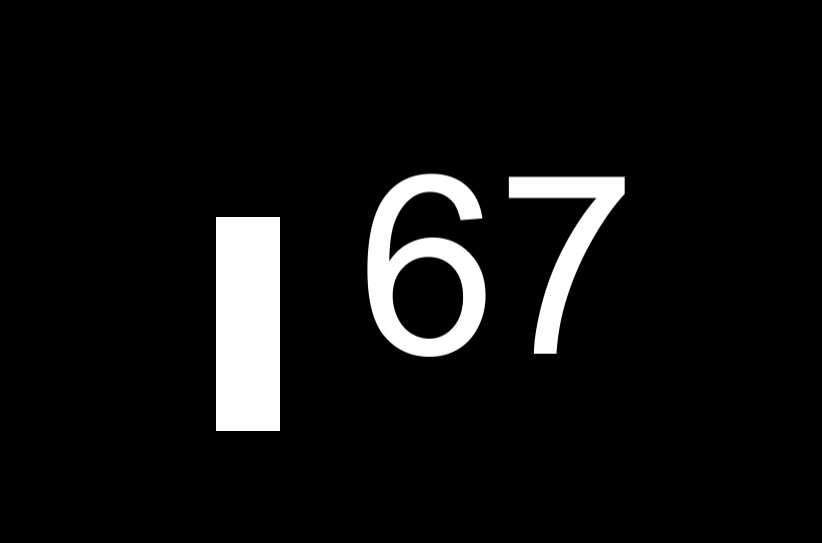}\label{fig:visReward}}
	\hspace{0.01\textwidth}
	\subfigure[]{\includegraphics[width=0.24\textwidth]{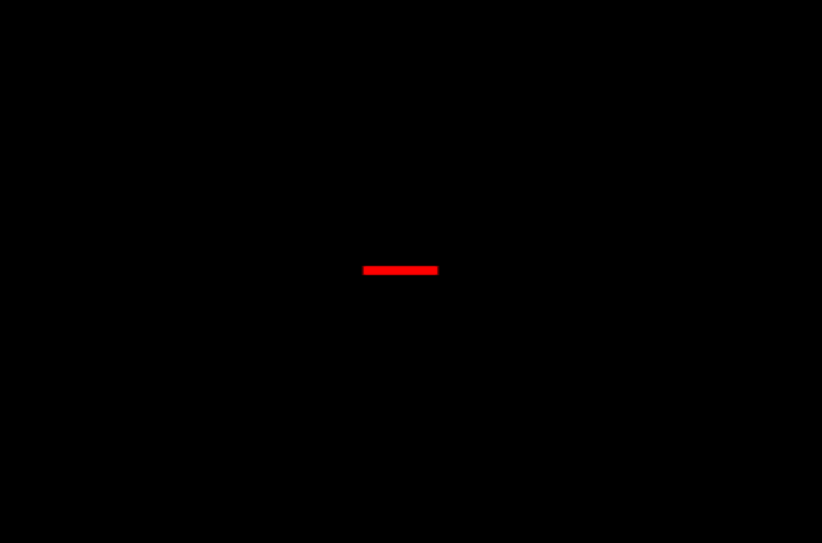}\label{fig:visFail}}
	\hspace{0.01\textwidth}
	\subfigure[]{\includegraphics[width=0.24\textwidth]{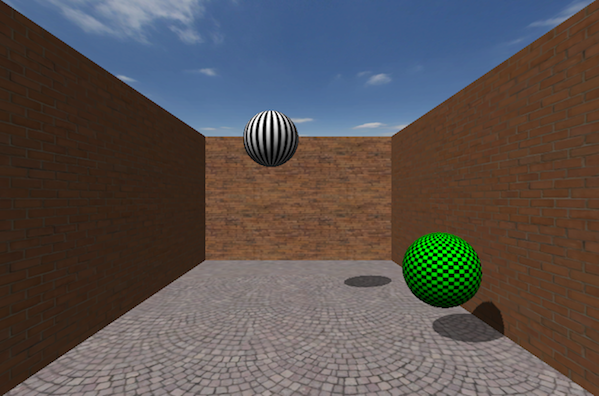}\label{fig:visReturn}}
    \caption{Presented visual feedback. (a) Baseline phase (5 sec): blank scene; subject is instructed to relax. (b) Planning phase (2.5--4 sec): white sphere represents the end-effector, and yellow sphere represents the target position; subject is instructed to plan their movement but to not yet move. (c) Go phase (max 10 sec): target's color switches to green; subject starts movement. (d) The rating, shown for two seconds as a number and by a vertical bar (height reflects value). (e) Feedback shown for failed or aborted trials. (f) Return phase: starting position is shown in green; subject is instructed to return.}\vspace{-0.1cm}
    \label{fig2}
\end{figure*}

The visual feedback consisted of a virtual reality scene in which the position of the end-effector and the target were displayed as colored and striped balls (see Fig.~\ref{fig2}). Because depth perception is a critical element in virtual three-dimensional reaching movements \cite{Howard:2012}, we used lifelike textures, lighting, shadows, and gloss. Each trial began with the \textit{baseline} phase, for which the subjects were instructed to do nothing and no feedback was shown. After 5 seconds, the \textit{planning} phase began, for which the subjects were instructed to plan the reaching movement but not yet move. Henceforth, the visualization displayed the positions of the end-effector and the current target; these positions were shown as spheres, colored in white and yellow for the end-effector and target position, respectively. The \textit{planning} phase lasted 2.5--4 seconds, chosen randomly from a uniform distribution. If the subjects moved the end-effector more than 4 cm from the starting position in this phase, the trial was aborted. Otherwise, it continued with the \textit{go} phase, visually indicated by switching the target's color from yellow to green. In the \textit{go} phase, subjects were instructed to reach for the target location to overlap the positions of the end-effector and target in the virtual scene. Movements were considered complete when the subjects moved the end-effector to within 3.5 cm of the target, or if the movement duration exceeded 10 seconds. After this phase, the calculated \textit{rating} was presented for 2 seconds. If subjects exceeded the 10 seconds time limit or performed a false start, a red dash was shown instead of the \textit{rating}. Finally, the \textit{return} phase instructed the subjects to return to the starting position. When the end-effector was moved to within 1 cm of the starting position, the next trial began at a new random target location.

In order to determine a range of reachable target positions while considering the physical differences in subjects, each subject was prompted before the experiment to move his arm to multiple comfortable positions in front of them by naturalistic reaching movements. We then calculated the minimum sphere that enclosed 90\% of all these reachable positions to which the subject moved his arm comfortably. From this sphere 50 equidistant target locations were chosen. In this study, the radii of the sphere varied across the subjects in a range of 9--14 cm.

\subsection{Quantification of Motor Performance}
\label{sec:narj}

We quantified subjects' motor performance at each trial by their normalized average rectified jerk (NARJ) \cite{Cozens:2003} as an index of motor skill. Jerk, the third derivative of position and rate of change in acceleration, is a key variable in motor learning \cite{Flash:1985} and is widely used in human-robot interaction \cite{Amirabdollahian:2002,Seki:2004}. Here, it is of particular interest since post-stroke motor recovery, as measured by patients' Fugl-Meyer scores \cite{Fugl:1974}, is highly correlated with change in jerk (i.e., movement smoothness) \cite{Krebs:1999,Rohrer:2002}. By choosing jerk to quantify motor performance, we claim that our empirical results with healthy subjects relate to clinically relevant measures of stroke recovery.

We computed the trial-specific NARJ value by, first, differentiating the end-effector's position in each of the three spatial dimensions, second, computing the magnitude of the resulting velocity vector, third, differentiating the velocity's magnitude twice, fourth, computing the absolute value of the resulting instantaneous jerk, fifth, taking the mean across all instantaneous jerks of one reaching movement, and, finally, normalizing (multiplying) the resulting average rectified jerk by the time to complete the movement to the power of three \cite{Cozens:2003}. Hence, we quantify trial movement performance by the mean jerk normalized by the time needed to complete the reaching movement. In each trial's \textit{rating} phase, subjects were informed about their movement performance by inversely mapping the NARJ value to a 0--100 range; a higher score denoted a smoother movement (see Fig.~\ref{fig:visPaths}  for examples of smooth and non-smooth trajectories).

\subsection{EEG Acquisition}

Throughout the experiments, a 120-channel EEG was recorded at a sampling rate of 500 Hz, using active EEG electrodes and BrainAmp DC amplifiers (BrainProducts, Gilching, Germany). Electrodes were placed according to the five percent electrode system for high-resolution EEG \cite{Oostenveld:2001}, using TPP9h location for the initial reference electrode. All data were re-referenced to a common average reference offline. EEG signals were acquired from the amplifier using BCI2000 and its extension BCPy2000 \cite{Schalk:2004}. Before and after the visuomotor learning task, resting-state EEG was recorded for five minutes where subjects were instructed to relax with eyes open, looking at a fixation cross displayed in the middle of the screen, as a baseline condition for the visuomotor learning task that involved visual processing \cite{Barry:2007}.

\subsection{EEG Data Analysis and Feature Extraction}

Training a prediction model that relates resting-state EEG data to subjects' motor skills is challenging because of the high-dimensional feature space (120 EEG channels times 150,000 data points per channel) with only a small number of samples to learn from (37 subjects). To avoid overfitting, an informative feature space was obtained by transforming the EEG data into a small number of relevant features. This was achieved by reducing the dimensionality of EEG data both in the spatial and temporal domains. For dimensionality reduction in the spatial domain, we used a group-level independent component analysis (ICA) procedure and selected a few independent components (ICs) that represent cortical processes consistently found across all 37 subjects. For temporal dimensionality reduction, we transformed the five-minute resting-state data of every IC into the spectral domain and then computed their mean bandpower in the $\alpha$-band (8--14 Hz), relying on a variety of studies on relevance of $\alpha$-rhythms in visuomotor learning \cite{Meyer:2014,Manuel:2018,Rilk:2011,Sigala:2014}.

To separate the data into group-wise statistically independent components, we first clipped the data to the five-minute resting state phases prior to the experiment, and then used a third-order Butterworth high-pass filter on the raw data at 3 Hz. We then pooled the resting-state data of all subjects and reduced them to 64 principal components before performing group-wise ICA. We applied a second-order blind identification algorithm (SOBI) to compute the ICs \cite{Belouchrani:1997}. We inspected each IC manually and rejected those ICs as non-cortical for which at least one of the following criteria applied \cite{Jung:2000,Delorme:2012}: (1) The spectrum did not show the 1$/$f-behavior typical of a cortical source. (2) The topography did not show a dipolar pattern. (3) The time-series appeared to be contaminated by detectable eye blinks or other noise sources such as 50 Hz line noise. The remaining six IC topographies are shown in the left column of Fig.~\ref{fig3}.

To perform temporal dimensionality reduction, for each IC's time series, we computed the mean $\alpha$-power in pre-experiment resting-state phases. This was performed by, first, subtracting the time-series' mean, second, applying a Hann window spanning the whole five-minute resting phase, third, applying a Fourier transform, fourth, computing the logarithm of the absolute values of the Fourier coefficients, and, fifth, taking the average over all log-bandpowers in the $\alpha$-range. For each subject, this resulted in a total of six resting-state $\alpha$-powers, one for each IC, which constituted our neural feature space. We also repeated the same procedure for the post-experiment resting-state data for later analyses.

\subsection{Predicting Visuomotor Learning Skill From Resting EEG}

\subsubsection{Visuomotor Learning Skill Performance Measures}

Despite the fact that several studies use parameters of exponential or power law fits to kinematic performance data in order to characterize learning, it is known that multiple independent processes contribute to motor skill acquisition with different time scales \cite{Smith:2006}, which leads to insufficient quantification of motor performance. Therefore we compute individual measures of motor skill variability under no {\it a priori} assumptions about the shape of the performance data, where we dissociate motor skill from learning rate. Since we are interested in an overall measure to quantify motor skill for each subject, rather than trial-to-trial changes in movement smoothness, we computed the average NARJ values across all 100 reaching movements as individual measures of \textit{visuomotor learning skill}. We discarded trials that were false starts, in which the target was not reached within 10 seconds, and the trials in which the NARJ value exceeded three standard deviations of each subject's mean NARJ across all reaching movements. For each subject, this procedure resulted in a single measure of visuomotor learning skill (i.e., average movement smoothness) extracted from subjects' kinematic data.

\begin{figure}[t!]
    \centering
    \includegraphics[width=0.48\textwidth]{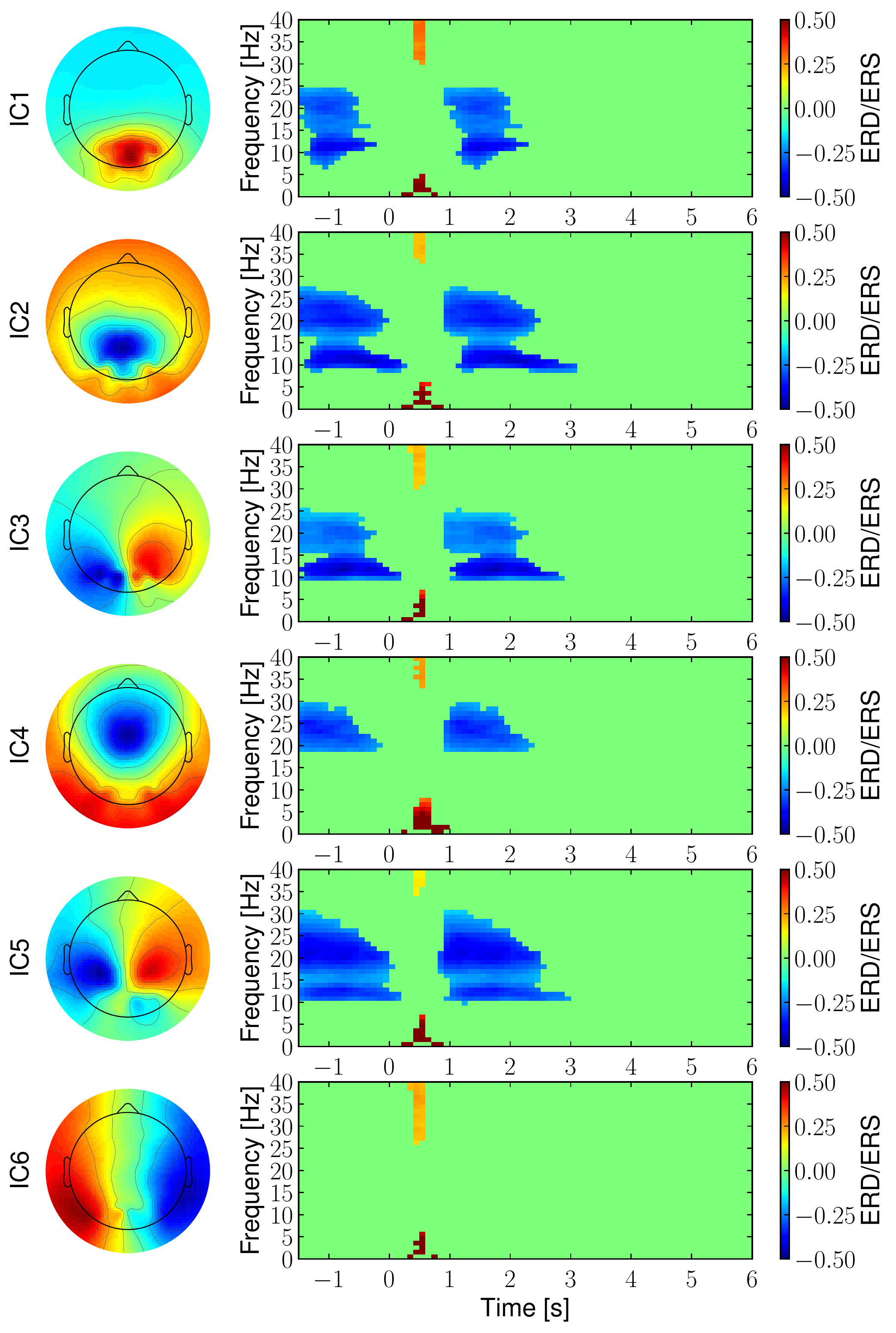}
    \caption{First column: Topographies of the six ICs used in the regression model. Second column: Group-average ERD/ERS patterns of ICs, time-locked to the instruction to reach to a target ($t=0$). Values on the x-axis indicate the end-points of the one-second time windows where bandpowers are computed. Non-significant ERD/ERS have been set to zero (FDR=5\%).}\vspace{-0.2cm}
    \label{fig3}
\end{figure}

\subsubsection{Training a Neural Prediction Model}

As an {\it a priori} outlier rejection criterion, we discarded subjects for which either any kinematic performance measure or neural feature value exceeded three standard deviations of the median across subjects \cite{Rousseeuw:2005}. Using the remaining subjects, we employed a leave-one-subject-out cross-validation procedure to learn a linear prediction model from each subject's pre-experiment configuration of brain rhythms to their visuomotor learning skill. Specifically, we first put one subject aside and then learned a multivariate linear least-squares regression model that explains subjects' mean NARJ values as a function of their six pre-experiment resting-state IC $\alpha$-powers. We then predicted the mean NARJ value of the held-out subject and re-iterated this procedure for every subject. We quantified the prediction performance by calculating the Pearson's correlation coefficient between the observed and the predicted mean NARJ values. To test the null-hypothesis of zero correlation, we randomly permuted the assignment of visuomotor learning skill measures to neural features across subjects 10,000 times and estimated the frequency at which the prediction model achieved a higher correlation coefficient than with the true assignment of brain rhythms to visuomotor learning skill measures (one-sided test). We rejected the null-hypothesis for a $p$-value below $\alpha = 0.05$.

To identify a final global prediction model defined by a global configuration of the brain rhythms used as neural features, we averaged the weights for each IC $\alpha$-power in the linear regression model across all cross-validation folds. To compare the predictive powers of individual brain rhythms within this configuration, we repeated the same prediction procedure using only one IC $\alpha$-power at a time.

\subsection{Neural Dynamics During Motor Planning and Execution}

Movement planning and execution is accompanied by a reduction in bandpower, known as event-related desynchronization (ERD), in the $\alpha/\mu$- (8--14 Hz) and $\beta$-ranges (20--30 Hz) over sensorimotor areas \cite{Pfurtscheller:1999}. To investigate which ICs are actively involved in motor planning and execution, we computed each IC's ERD in a range from zero to 40 Hz, time-locked to the instruction to initiate a reach. For each IC, subject, and trial, we first computed bandpower in bins of one Hz width using a window of one second length and a step-size of 100 ms. To obtain the individual ICs' ERD, we then averaged each IC's bandpowers across trials and subjects and subtracted, from each window, the mean bandpower across all windows of the last 2.5 seconds of \textit{baseline} phase (-2.5 seconds to the start of the \textit{planning} phase). To investigate which windows and frequency bins display a statistically significant ERD, we again employed a permutation test. We randomly permuted the order of all time windows 10,000 times and estimated the frequency at which the absolute changes in ERD exceeded those observed with the ordering of windows intact. We used a two-sided test to also test for potential event-related synchronization (ERS). We rejected the null-hypothesis for a bin and time window at a false discovery rate (FDR) of $\alpha_{\text{FDR}} = 0.05$ \cite{Benjamini:1995}.

Furthermore, to study the ERD of the global configuration of ICs' over the course of motor planning and execution, we first computed the dot product between the weights of the global linear regression model and the ICs' bandpowers, and then repeated the ERD computation on this global configuration of brain rhythms.

\begin{figure}[b!]
    \centering\hspace{-0.4cm}
    \includegraphics[width=0.44\textwidth]{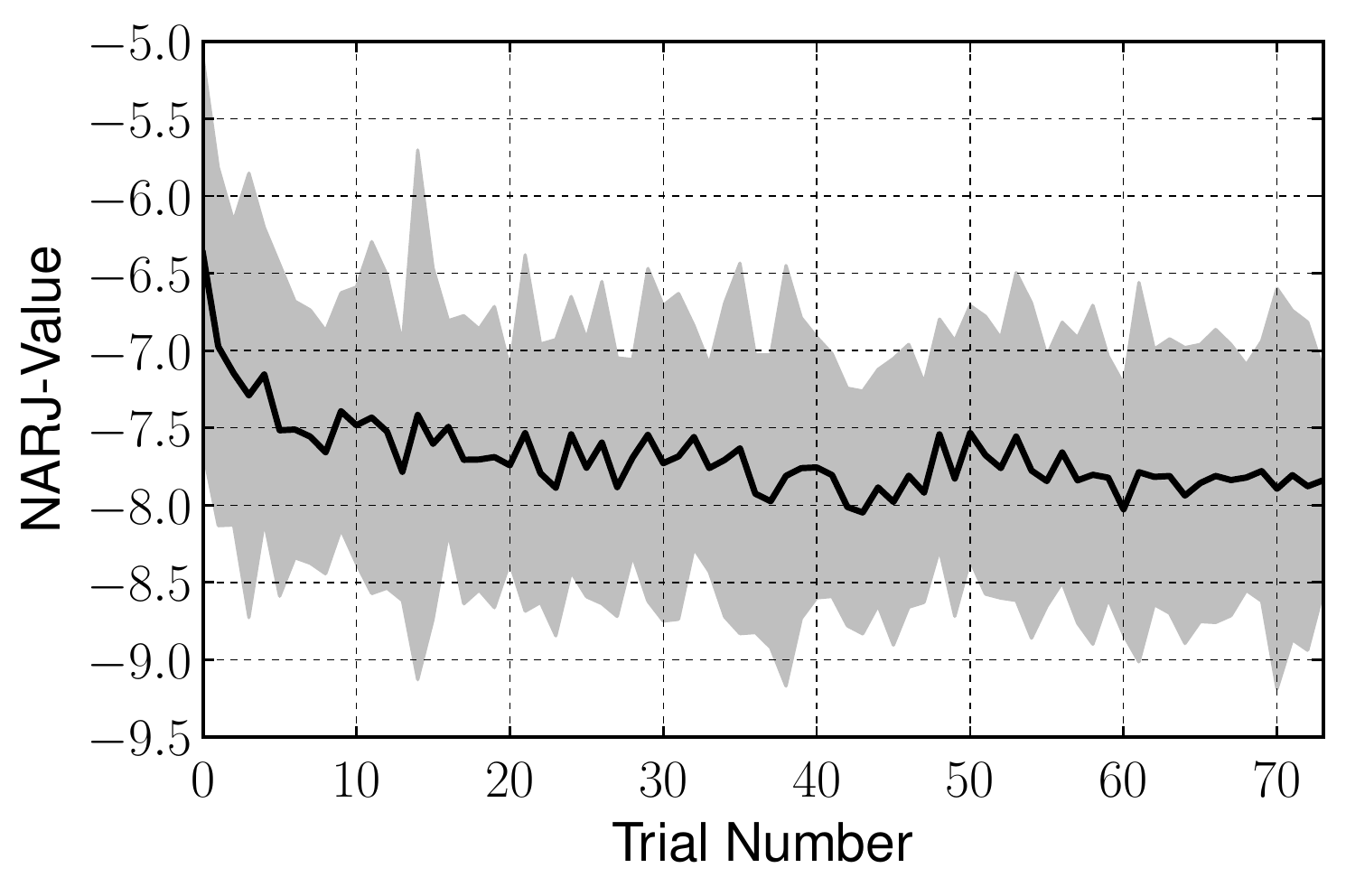}
    \caption{Group-median of NARJ-values across trials. Shaded areas represent plus/minus one standard deviation.}
    \label{fig4}
\end{figure}

\subsection{Neural Dynamics After Visuomotor Learning}

To test whether the brain rhythms changed over the experiment with learning on a single-IC level, we computed the difference in $\alpha$-power of each IC between the pre- and post-experiment resting-state baselines and used a Wilcoxon signed-rank test to test the null-hypothesis of zero median.

We also tested whether the predictive power of the global configuration of brain rhythms has changed between the pre- and post-experiment resting-state baselines. We did this by, first, computing the difference in predicted mean NARJ before and after the learning task (i.e., using the learned global prediction model on the pre- and post-experiment resting-state data), and, second, performing a paired permutation test, in which we randomly permuted the assignment of IC $\alpha$-powers to either pre- or post-experiment resting-state 10,000 times. We estimated the $p$-value for rejecting the null-hypothesis of no change in this global configuration as the frequency at which the resulting absolute difference in predicted mean NARJ exceeded the one obtained with intact ordering of $\alpha$-powers (two-sided test).

\section{Results}

\subsection{Visuomotor Learning Skill Varies Across Subjects}

We discarded on average $3.54\pm3.83$ trials per subject due to false starts and $0.76\pm1.45$ trials because the target was not reached within the 10 seconds time limit. Another $3.11\pm0.99$ trials were discarded on average per subject due to excessively large NARJ values. This resulted in a minimum of 74 trials, out of 100, available per subject. Fig.~\ref{fig4} displays the group-median NARJ values versus trial numbers. We observe an initially steep learning curve for ten trials, which is followed by a more gradual increase in movement smoothness. After about 40 trials, no further improvement in smoothness is apparent as performance reaches a steady-state. We observed that these data points follow a power law with a significant correlation between an exponential fit and NARJ-values ($\rho = 0.84,\ p < 10^{-3}$, where the $p$-value is estimated by a random permutation test for the null hypothesis of having no correlation). This shows a consistency suggested by law of practice \cite{Newell:1981}.

\begin{figure}[b!]
    \centering\hspace{-0.3cm}
    \includegraphics[width=0.44\textwidth]{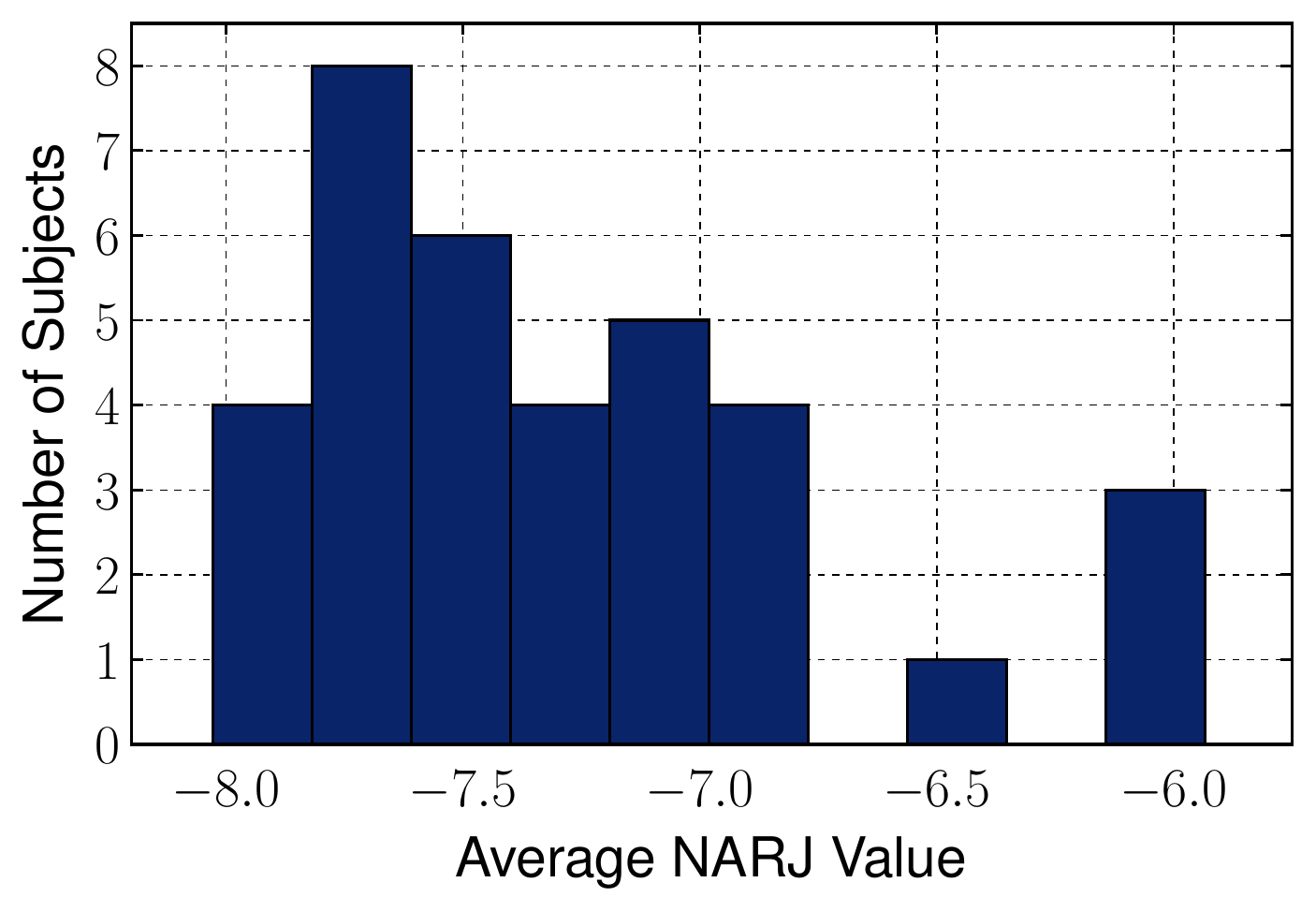}
    \caption{Histogram of average NARJ values across participants. Lower values represent more smooth movements.}
  \label{fig5}
\end{figure}

The histogram of subject-average NARJ values across 100 trials as individual visuomotor learning skill measures are shown in Fig.~\ref{fig5}. Average NARJ values for most subjects cluster around a value of $-7.75$, with a long tail of the distribution towards less smooth movements. These results demonstrate that the experimental paradigm establishes the presence of visuomotor learning, and captures substantial variations in visuomotor learning skill across subjects.

\subsection{Global Configuration of Resting-State $\alpha$-Rhythms Predicts Across-Subjects Variations in Movement Smoothness}

Topographies of the obtained six cortical ICs are shown in the left column of Fig.~\ref{fig3}. All ICs exhibit a prominent dipolar pattern. ICs 1--3 depict a parietal focus, where IC 1 is likely to represent a parieto-occipital source. IC 4 shows a fronto-central pattern, which is likely to be associated with cortical processes linked to fronto-parietal attention networks together with ICs 1--3 \cite{Bressler:2010}. IC 5 exhibits a distinct focus over sensorimotor areas, indicating a central role in motor execution \cite{Pfurtscheller:1999}. The widespread topography of IC 6 is likely to represent a deep sub-cortical source.

\begin{figure}[t!]
    \centering\hspace{-0.4cm}
    \includegraphics[width=0.43\textwidth]{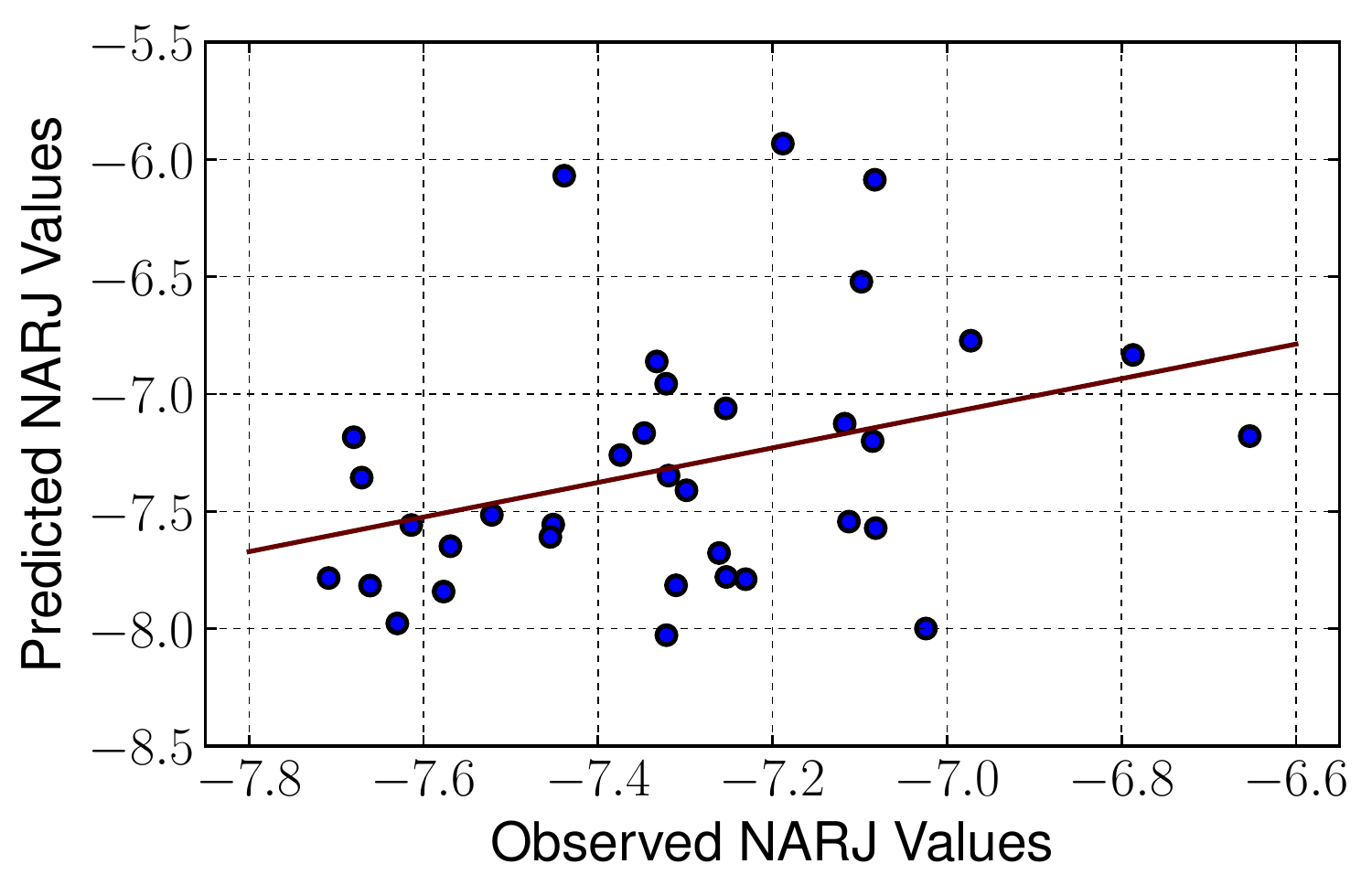}
    \caption{The global configuration of resting-state $\alpha$-rhythms predicts average movement smoothness (NARJ) in a visuomotor learning task ($\rho = 0.34,\ p = 0.02$). One dot represents one subject.}\vspace{-0.1cm}
    \label{fig6}
\end{figure}

\begin{figure}[t!]
    \centering\hspace{-0.4cm}
    \includegraphics[width=0.43\textwidth]{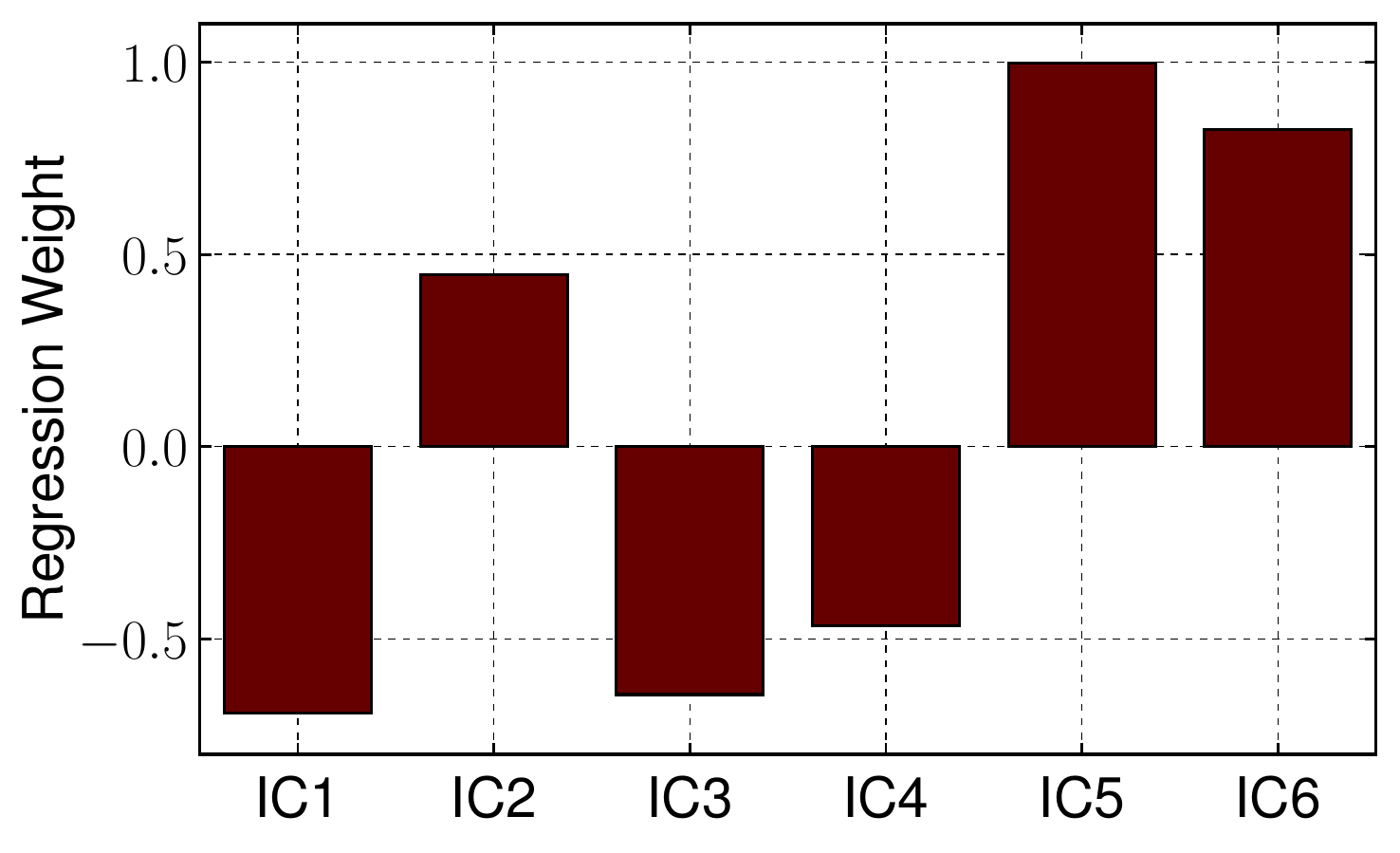}
    \caption{Weights of the global multivariate linear regression model for each IC $\alpha$-power, obtained by averaging across cross-validation folds.}
    \label{fig7}
\end{figure}

One subject was excluded from prediction model training due to excessively poor motor performance, and one subject was discarded due to excessively high brain rhythms and outlier neural features. For the remaining 35 subjects, the global configuration of resting $\alpha$-powers of the six ICs significantly predicted their visuomotor learning skill ($\rho=0.34$, $p = 0.02$, see Fig.~\ref{fig6}). Fig.~\ref{fig7} displays the regression weights that were assigned to each IC by the prediction model. We note that the model assigned a non-zero weight to every IC. We also found none of the ICs to predict subjects' motor skill on their own $\left(p_{\text{IC1-6}} = \{0.55,\ 0.51,\ 0.92,\ 0.47,\ 0.17,\ 0.16\}\right)$. This indicates that the neural signature of motor skill is found in the relation of brain rhythms, but not in the amplitude of individual brain rhythms.

\subsection{EEG Dynamics During Motor Planning and Execution}

The ERD patterns of all ICs, time-locked to the instruction to initiate a reach, are displayed in the right column of Fig.~\ref{fig3}. As expected from their putative roles in motor planning and execution, ICs 1--3 and IC 5 display an ERD in the $\alpha\ (\mu)$- (8--14 Hz) and in the $\beta$-range (20--30 Hz) during motor planning (-1.5--0 seconds) and again during motor execution (1--3 seconds). IC 4 only exhibits an ERD in the $\beta$- but not in the $\alpha$-band. IC 6 shows no task-related modulation, potentially due to representing sub-cortical processes. The ERS in very low and in high frequencies, that is visible in all ICs, coincides with the initiation of a movement at around 200 ms and is thus very likely to be a movement artifact.

\begin{figure}[t!]
    \centering
    \includegraphics[width=0.45\textwidth]{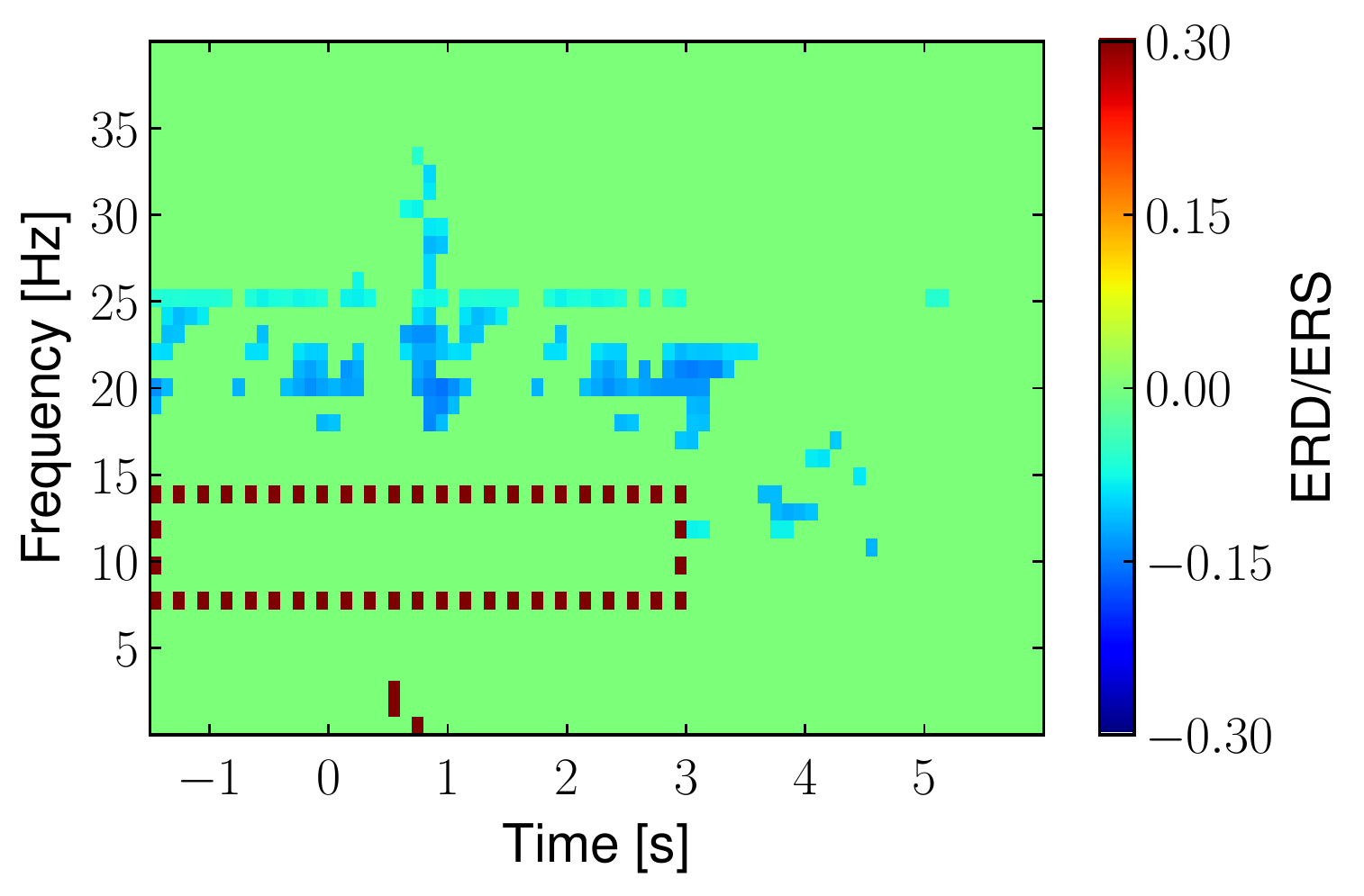}
    \caption{Group-average ERD/ERS of the global configuration of brain rhythms predictive of visuomotor learning skill, time-locked to the instruction to reach a target ($t=0$). Values on the x-axis indicate the end-points of the one-second time windows where bandpowers are computed. Non-significant ERD/ERS have been set to zero (FDR=5\%). The dotted red lines enclose the time-frequency range of $\alpha$-ERDs of individual ICs (cf.~Fig.~\ref{fig3}).}\vspace{-0.1cm}
    \label{fig8}
\end{figure}

\subsection{Neural Signature of Motor Skill Is Independent of Motor Planning and Motor Execution}

Fig.~\ref{fig3} shows clear ERD patterns of ICs 1--5 in the $\alpha$-range during motor planning and execution (-1.5--3 seconds). Similarly, any random linear combination of these brain rhythms is also likely to exhibit an ERD/ERS in the $\alpha$-range. However, the global configuration of $\alpha$-rhythms that predicts subjects' motor skill does not show any task-related modulation in the specific frequency- (8--14 Hz) and time-range (-1.5--3 seconds) in which the individual ICs exhibit task-related ERDs (cf.~Fig.~\ref{fig8}). In mathematical terms, this is equivalent to the global configuration (the neural signature of motor skill) being orthogonal to task-related $\alpha$-power changes. We highlight this finding for motor rehabilitation, since it enables us to dissociate the neural signature of motor skill from ERDs related to motor planning and execution.

\subsection{Neural Signature of Motor Skill Is Not Altered After Visuomotor Learning}

\begin{figure}[t!]
    \centering\hspace{-0.4cm}
    \includegraphics[width=0.43\textwidth]{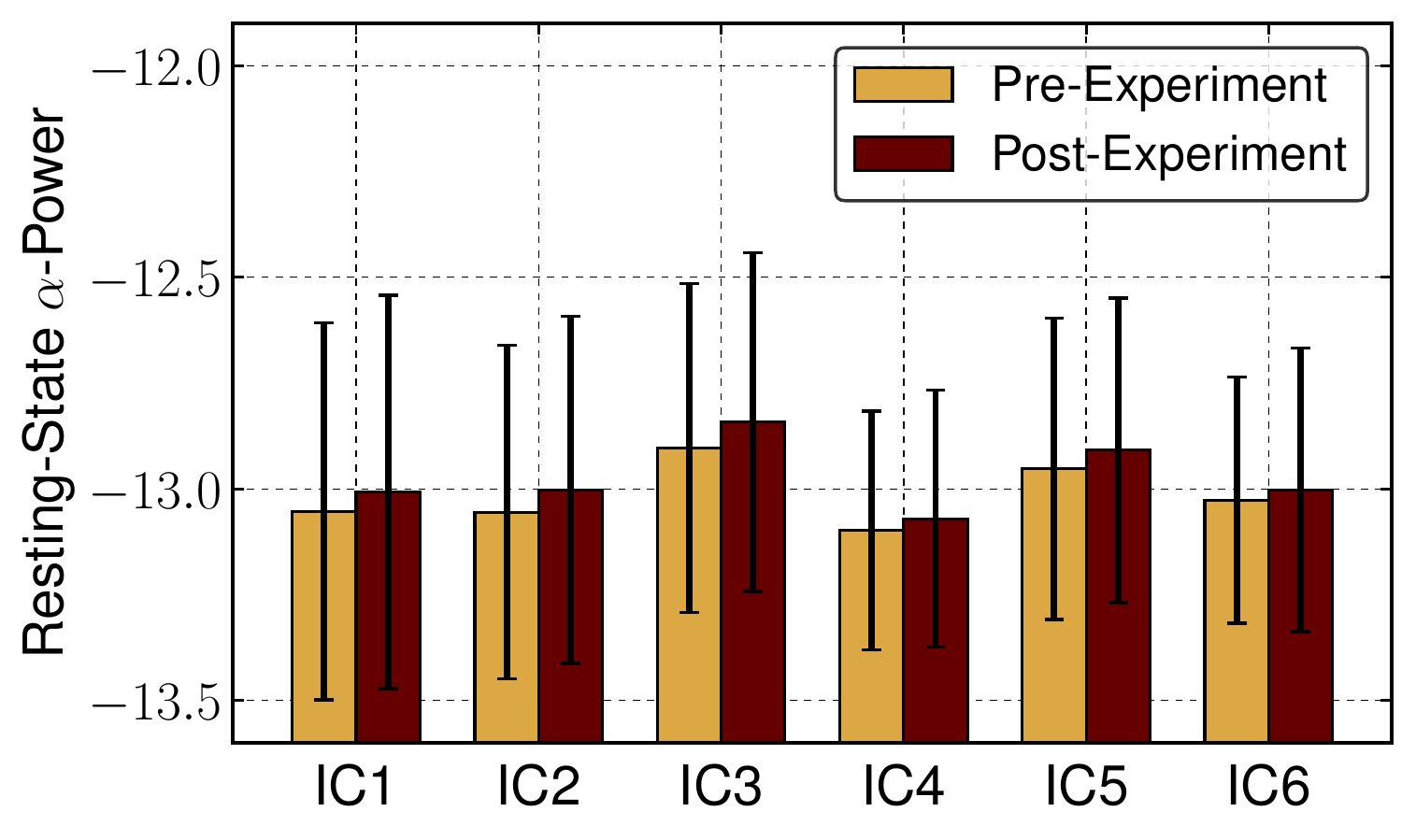}
    \caption{Pre- and post-experiment values of resting-state $\alpha$-power. Error bars indicate the standard deviation across subjects.}\vspace{-0.1cm}
    \label{fig9}
\end{figure}

We observed a statistically significant change in $\alpha$-power of ICs between pre- and post-experiment resting-state baseline recordings ($p = 0.03$). As shown in Fig.~\ref{fig9}, the $\alpha$-powers of all six ICs increased between the pre- and the post-experiment resting-state data. However, we did not find a change in the predictive power of the global configuration of brain rhythms between the pre- and post-experiment resting-states ($p = 0.99$). This indicates that the neural signature of motor skill is not altered by visuomotor learning itself.

\section{Discussion}

We have presented empirical evidence that the global configuration of $\alpha$-rhythms in the resting brain contains a neural signature of subjects' motor skills in visuomotor learning. We found this neural signature to be orthogonal to (independent of) task- as well as to learning-related changes in $\alpha$-rhythms. Because it is {\it a priori} unlikely that any random combination of $\alpha$-rhythms across multiple brain areas is orthogonal to their task-related changes, we interpret this dissociation of the neural signature of motor skill and brain rhythms involved in motor execution and planning as an organizing principle of the human brain. Specifically, our finding implies that task-induced changes in $\alpha$-rhythms are globally coordinated to preserve the neural signature of motor skill, and this configuration is more likely to reflect a fundamental organization of a subject's brain rhythms rather than a momentary state-of-mind. Furthermore, our observations on post-learning alterations of individual resting brain rhythms is consistent with previous evidence \cite{Vahdat:2011,Moisello:2013,Ozdenizci:2017b}, while the predictive neural signature of motor skill is found to be preserved. Similarly, the brain rhythms involved in the neural signature of motor skill (i.e., ICs) were found to be consistent with previous work on EEG identifiers of motor adaptation learning performance \cite{Ozdenizci:2017}, as well as various other neuroimaging studies on the associated roles of the observed distinct cortical sources involved jointly in human motor learning skills \cite{Classen:1998,Luu:2004,Culham:2006}.

Statistical analyses with other frequency band neural prediction models (i.e., $\theta$-band (4--7 Hz), $\beta$-band (15--30 Hz), $\gamma$-band (55--85 Hz)) did not yield significant results. Furthermore, due to the small number of subject data in comparison to high dimensionality of possible neural feature spaces (i.e., combinations of various frequency bands and cortical sources), we were not able to investigate a wider range of neural signatures but focused on the $\alpha$-range which was shown to be related with visuomotor behavior. Yet, we believe that performing a similar but broader series of analyses with more subject data may yield stronger results.

We consider healthy subjects' and stroke patients' motor skills to lie on a continuum, with severely impaired patients at the lower end and healthy athletes at the upper end of the spectrum \cite{Rohrer:2002,Schmidt:2018}. We hypothesize, first, that such neural signatures of motor skill correlate with the severity of motor deficits in stroke patients, and, second, that teaching stroke patients to reconfigure their neural signature of motor skill may support motor recovery. This reconfiguration may be achieved by BCI-assisted rehabilitation robots that monitor a patient's brain rhythms real-time during rehabilitation exercises, and whenever they detect a change that is likely to lead to better performance, reward this change by providing active movement support \cite{Gomez:2011,Ramos:2012,Sarac:2013,Ang:2014,Ang:2015}. In this way, patients may be able to learn how to generate a configuration of LSCNs that supports motor recovery. Nevertheless, feasibility of such a neurorehabilitation approach rests on open questions. The first question is whether the results obtained with healthy subjects in this study translate to stroke patients. The second question is whether the visuomotor learning task used in this study generalizes to motor tasks that are relevant to stroke patients' activities of daily living. Finally, it remains an open question whether patients are able to reconfigure their neural signature of motor skill by neurofeedback \cite{Ozdenizci:2014}, and if this can be generalized to stroke patients individually \cite{Mastakouri:2017}.


\section*{Acknowledgment}

This work was partially supported by the Scientific and Technological Research Council of Turkey by a graduate fellowship, and by Sabanc{\i} University Internal Research Grant Program under IACF-11-00889.



\end{document}